\documentclass[]{emulateapj}

\shorttitle{CO(1-0) in GRB Host 080517}
\shortauthors{Stanway et al.}

\begin{document}

\title{A Detection of Molecular Gas Emission in the Host Galaxy of GRB\,080517}

\author{E.~R.~Stanway\altaffilmark{1} and A.~J.~Levan}
\affil{Department of Physics, University of Warwick, Gibbet Hill Road, Coventry, CV4 7AL, UK}

\author{N.~R.~Tanvir and K.~Wiersema}
\affil{Department of Physics and Astronomy, University of Leicester, University Road, Leicester LE1 7RH, UK}

\and

\author{T.~P.~R.~van der Laan} 
\affil{Institute de Radioastronomie Millimetrique (IRAM), 300 Rue de la Piscine, 38406 St. Martin d’Heres, Grenoble, France}

\altaffiltext{1}{email:e.r.stanway@warwick.ac.uk \email{e.r.stanway@warwick.ac.uk}}

\begin{abstract}
We have observed the host galaxy of the low redshift, low luminosity
GRB\,080517 at 105.8\,GHz using the IRAM Plateau de Bure
interferometer. We detect an emission line with integrated flux
$S\Delta\nu = 0.39\pm0.05$\,Jy\,km\,s$^{-1}$ -- consistent both
spatially and in velocity with identification as the J=1-0 rotational
transition of carbon monoxide (CO) at the host galaxy redshift. This represents only the third long GRB
host galaxy with molecular gas detected in emission. The inferred
molecular gas mass, $M_\mathrm{H_2}\sim6.3\times10^8$\,$M_\odot$,
implies a gas consumption timescale of $\sim$40\,Myr if star formation
continues at its current rate. Similar short timescales appear
characteristic of the long GRB population with CO observations to
date, suggesting that the gamma-ray burst in these sources occurs towards the end
of their star formation episode.
\end{abstract}


\section{Introduction}\label{sec:intro}
Long Gamma-Ray Burst (GRB) host galaxies are selected out by the
explosion that is believed to mark the end of the life of a massive
star \citep[e.g.][]{2006ApJ...637..914W}. Given the short lifetime of
such stars, it is unsurprising that the vast majority of host galaxies
to date have been identified as star forming
\citep[e.g.][]{2009ApJ...691..182S,2010MNRAS.tmp..479S} and are
predicted to be rich in molecular gas
\citep{2002ApJ...569..780D,2000ApJ...532..273D}. The
absence of absorption signatures in the afterglows of bursts,
illuminating molecular gas along their line of sight out of their host
galaxy, has thus caused some concern. Initial investigations
\citep[e.g.][]{2004A&A...419..927V,2009A&A...506..661L} revealed a
shortage of detections in the H$_2$ Lyman-Werner bands relative to
those expected. Unusually low metallicity, low dust content
conditions, preventing the formation of molecules on the surface of
dust grains, have been invoked as a possible explanation
\citep[][]{2009A&A...506..661L,2007ApJ...668..667T}.  While a handful
of GRBs have since been identified with molecular absorption
signatures in their afterglows
\citep{2009ApJ...691L..27P,2013A&A...557A..18K,2014A&A...564A..38D,2014arXiv1409.6315F},
these observations are difficult and require sensitive spectroscopy
probing the rest-frame ultraviolet.

In an alternative approach, a number of GRB host galaxies have been
targeted for observations of direct emission from carbon monoxide in
the far-infrared rotational transitions. Assuming a conversion factor,
derived from local resolved star formation, this can be used as a
proxy for the presence of molecular hydrogen \citep[see][and
  references therein]{2005ARA&A..43..677S}. Early observations
targetting the host of low redshift GRBs\,980425
\citep[$z=0.0085$]{2007PASJ...59...67H} and 030329
\citep[$z=0.17$]{2007ApJ...659.1431E} yielded non-detections, as did
observations of the higher redshift host galaxies of GRBs\,000418
\citep[$z=1.1$]{2011ApJ...738...33H} and 090423
\citep[$z=8.2$]{2011MNRAS.410.1496S}. It is only recently that
observations with the highly sensitive Atacama Large Millimeter Array (ALMA) secured a detection of
carbon monoxide emission in the hosts of GRBs\,020819B and 051022 \citep[$z=0.41$ and
  0.81]{2014Natur.510..247H}.

We have recently identified the host of GRB\,080517 as an example of
an unusually low redshift ($z=0.089\pm0.003$) GRB host galaxy, with evidence for a mature stellar
population underlying an ongoing star formation episode \citep{2014arXiv1409.5791S}. The host
galaxy is accompanied by a neighbouring galaxy, sufficiently close in velocity and projected distance
to plausibly be interacting with it.
In this Letter we present observations targetting the J=1-0 rotational transition of carbon monoxide
in this system\footnote{Based on observations carried out with the IRAM Plateau de Bure Interferometer. IRAM is supported by INSU/CNRS (France), MPG (Germany) and IGN (Spain).}.


\section{Observations}\label{sec:obs}

Observations, centered on the X-ray location of GRB\,080517 (06$^h$
48$^m$ 58.03$^s$ +50$^\circ$ 44$'$ 07.7$''$, 90\% error radius
1.5$''$), were undertaken at the IRAM Plateau de Bure Interferometer
(PdBI) on 2014 Jun 09 and 2014 Jun 11. The WideX correlator was used
to probe a 3.6\,GHz bandwidth centered at 105.8\,GHz, the expected
frequency of the CO(1-0) transition at the target redshift, at 2\,MHz
resolution. Data were collected by observatory staff and scheduled
across hour angles ranging from -4.1 to 1.2h to ensure good $uv$-plane
coverage. A total integration time of 72 minutes was achieved on
target. The array was in the compact 5Dq configuration, with five
antennae in use. The resulting synthesized beam had a full-width at
half-maximum (FWHM) of $5.7\times4.4''$ and the half-power width of
the PdBI primary beam at this frequency is 48$''$. Bandpass and flux
calibration was performed on 3C454.3, yielding an estimated systematic
uncertainty $<$10\%. Phase calibration was performed with reference
observations of QSO B0714+457, which has a flux of 0.2\,Jy at these
frequencies, taken at regular intervals throughout the science
observing. Data were reduced and calibrated using standard tools in
the GILDAS data reduction package, and were imaged using natural
weighting.

\begin{figure*}
\epsscale{.99}
\plottwo{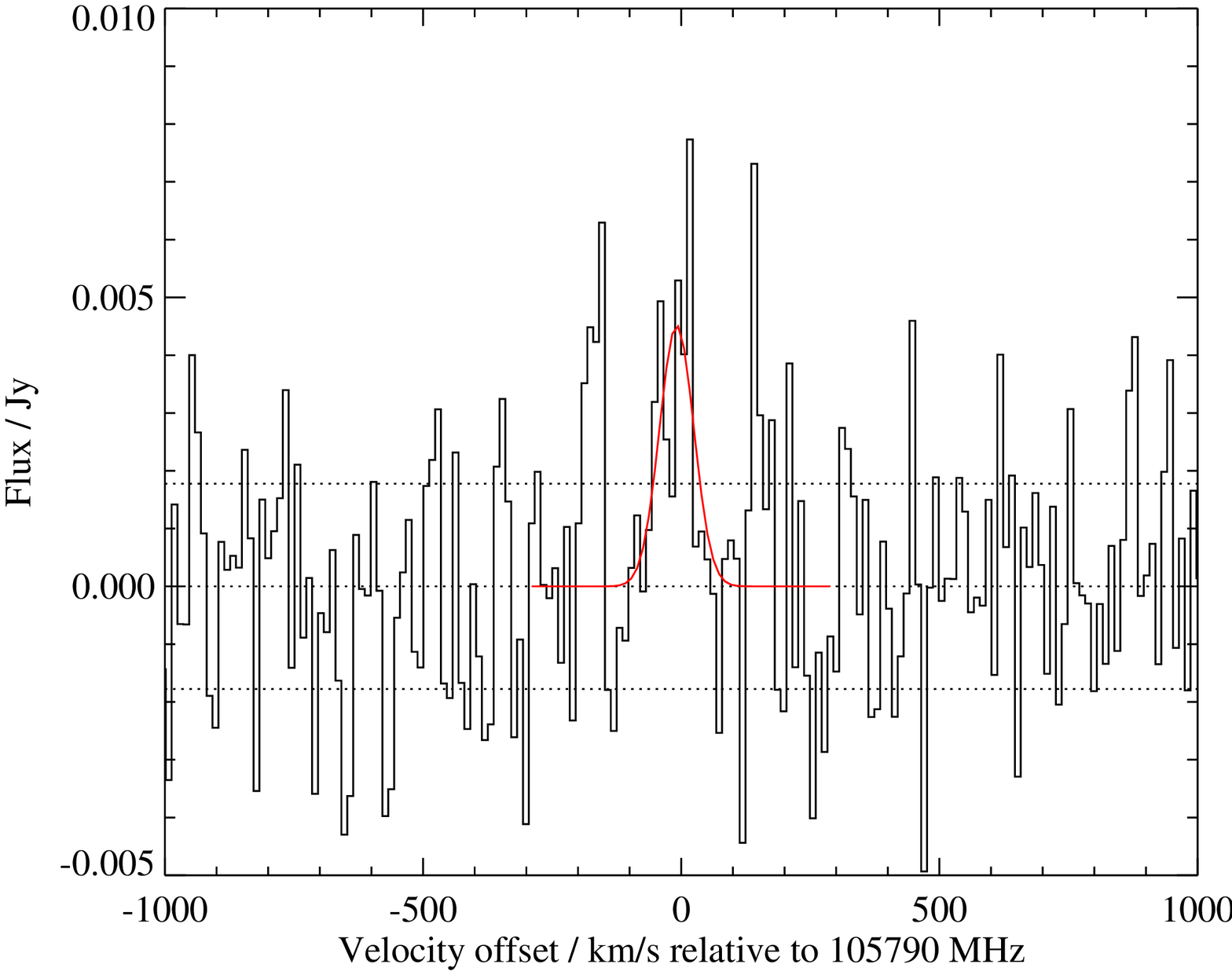}{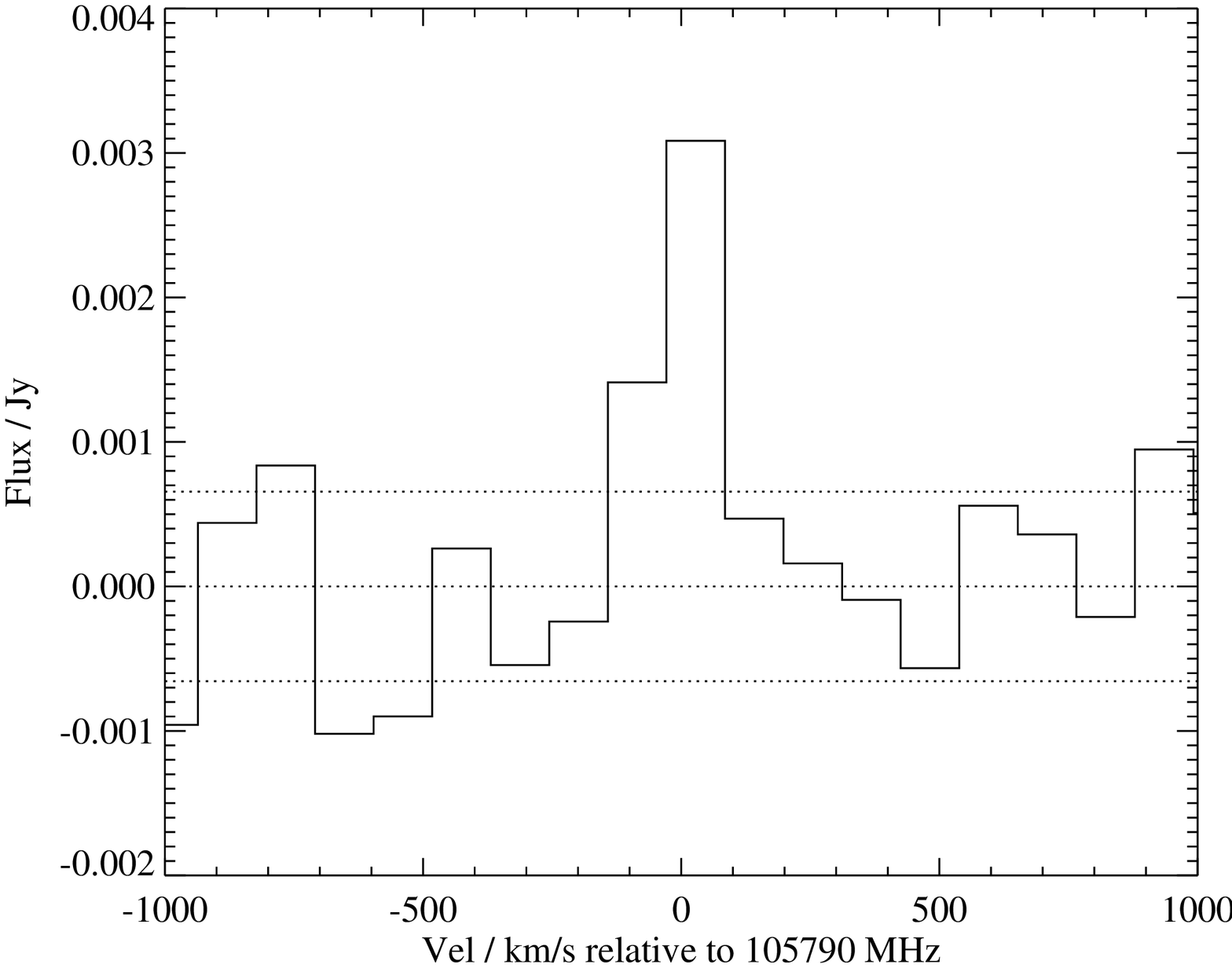}
\caption{The spectrum of GRB Host\,080517 at velocity resolutions of
  4\,MHz (left) and 40\,MHz (right). Spectral information is extracted
  from a spatial region corresponding to 1.5 times the synthesized
  beam, located at the position of the GRB\,080517 host galaxy. We
  measure a flux excess at the expected frequency of the CO\,(1-0)
  rotational transition. Dotted lines indicate the root-mean-square noise level. \label{fig:line_spec}}
\end{figure*}

A flux excess was observed in the resulting data cube at the pointing
centre and target frequency.  We extract spectral information from
a spatial region corresponding to 1.5 times the synthesized beam. The centre of this
aperture is located at
the optical position of the GRB\,080517 host galaxy (i.e. within 2$''$ of the
pointing centre). As figure \ref{fig:line_spec} illustrates, the
resulting spectrum shows a clear emission line centred at
105.790\,GHz. This corresponds to $z=0.08962\pm0.00003$ if the line is
identified, as we expect, as the CO J=1-0 rotational transition with a
rest-frame frequency of 115.271\,GHz.

\begin{figure}
\epsscale{1.0}
\plotone{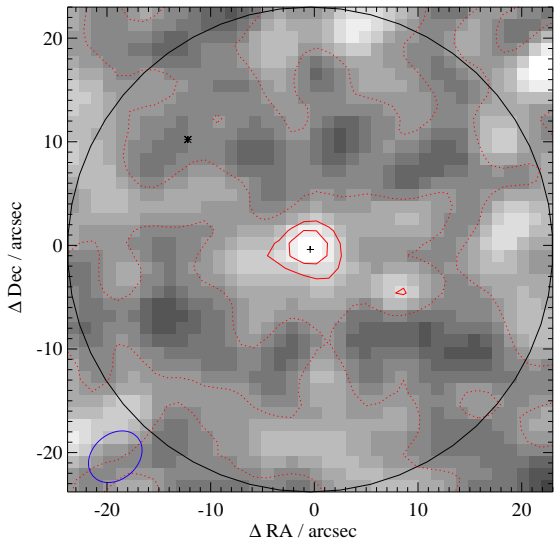}
\caption{The CO emission from GRB Host 080517, summed over a 40 MHz bandwidth centred on the peak of the emission line. A + symbol marks the position of the GRB host galaxy, while an asterisk indicates the position of the neighbouring galaxy. Contours indicate the zero flux level (dotted) and +2 and 3\,$\sigma$ (solid). The 1\,$\sigma$ noise at the pointing centre of this map is 0.3\,mJy\,beam$^{-1}$. There are no -2\,$\sigma$ peaks in the map. The large circle indicates the half-power extent of the PdBI primary beam. The ellipse at the lower left indicates the size of the synthesized beam.\label{fig:line_map}}
\end{figure}

We re-imaged the data, integrating those channels contributing to the
line, yielding a total bandwidth of 40\,MHz (113\,km\,s$^{-1}$). The resulting image
(figure \ref{fig:line_map}) reveals an isolated source located
coincident with the optical centre of the GRB host
galaxy. Deconvolution with the synthesized beam yields a marginally
resolved source with FWHM $2.0\times2.5''$, consistent with the
1.7$''$ Sersic radius measured in the optical (although with considerable uncertainty since
this source is well below the size of the synthesized beam).

The observed emission line has a peak flux 7.7$\pm$1.7\,mJy, measured
in a 4\,MHz channel. A gaussian fit to the observed line yields a
full-width at half maximum of 76.5\,km\,s$^{-1}$ and an integrated
line flux $S\Delta\nu = 0.39\pm0.05$\,Jy\,km\,s$^{-1}$ (a 7\,$\sigma$
detection). A possible second velocity component is observed at a
velocity offset of $\sim-170$\,km\,s$^{-1}$, but at too low a
significance to be considered a detection ($\sim$2\,$\sigma$). No
2.8\,mm continuum flux is detected from the GRB host galaxy to a
3$\sigma$ limit of 0.4\,mJy beam$^{-1}$ in a map constructed from line-free channels, with a
total bandwidth of 3.4\,GHz.

The neighbouring galaxy described in \citet{2014arXiv1409.5791S} is
offset from the GRB host by 16$''$ and $\Delta
v=576\pm155$\,km\,s$^{-1}$, allowing us to simultaneously constrain it
with this data. Neither line nor continuum flux is detected from the
neighbouring galaxy to the limits of the data. We note that one
component of the neighbour (Component B in Stanway et al 2014) is
coincident with a 2\,$\sigma$ peak in the continuum flux map,
accounting for primary beam attenuation at the galaxy location (a
factor of 0.65). However other peaks of comparable or greater
statistical significance have no optical-infrared counterpart,
suggesting that this may well be coincidence with correlated noise in
the map.

\section{Inferred Quantities}\label{sec:inf}

We convert the integrated line flux obtained above to a CO luminosity\footnote{adopting a standard H$_0=70$\,km\,s$^{-1}$\,Mpc$^{-1}$,
$\Omega_M$=0.3, $\Omega_\Lambda$=0.7 cosmology}. The resulting line
luminosity, L$'_\textrm{CO}=1.5\times10^8$\,K\,km\,s$^{-1}$\,pc$^{2}$,
corresponds to an inferred molecular gas mass, M$_\mathrm{H_2}
=\alpha$\,L$'_\textrm{CO}=6.3\times10^8$\,$M_\odot$ for the Galactic
conversion factor $\alpha=4.3$\,$M_\odot$\,/\,(K\,km\,s$^{-1}$\,pc$^{2}$) \citep{2013ARA&A..51..207B,2009ApJ...691L..27P}. This
amounts to only 15\% of stellar mass derived from spectral energy
distribution fitting \citep{2014arXiv1409.5791S} and potentially less
if the $\alpha=0.8$ conversion factor appropriate for densely star forming
systems such as ULIRGs is more appropriate.

Given a measured luminosity in the CO(1-0) line, we can estimate the
likely far-infrared luminosity, assuming that the host galaxy
follows the well-established \citep[see][and references therein]{2005ARA&A..43..677S} correlation between these
quantities. The inferred L$_\mathrm{FIR}=2\times10^{10}$\,L$_\odot$
would correspond to a predicted 105.8\,GHz continuum flux of just 8\,$\mu$Jy,
assuming a modified blackbody with dust temperature 35\,K and
emissivity index $\beta=2.0$, consistent with our non-detection in the
continuum. Given this modified blackbody spectrum, the predicted submillimeter flux at 850\,$\mu$m
(350\,GHz) would be $S_{850}=0.8$\,mJy, challenging but within reach of the
current generation of instruments. 

Interestingly, we can compare this prediction for the submillimeter
luminosity with estimates based on earlier
observations. The H$\alpha$ and 22$\mu$m-continuum derived star formation rate of
$\sim16$M$_\odot$\,yr$^{-1}$ \citep{2014arXiv1409.5791S} would lead to a predicted thermal
infrared luminosity L$_\mathrm{FIR}\sim10^{11}$\,L$_\odot$, while the
radio continuum emission detected at 4.5\,GHz would suggest a thermal
infrared luminosity L$_\mathrm{FIR}\sim6\times10^{9}$\,L$_\odot$ \citep[using conversions from][]{2012ARA&A..50..531K}. The
estimate from the carbon monoxide line emission is bracketed by these
alternate estimates. Given that each conversion from flux to star formation rate
is associated with a $\sim30$\% error, and that the scatter in the
L$'_{CO}$-L$\mathrm{FIR}$ relation is $\sim0.5$\,dex, these results
show broad agreement, producing a coherent picture of the emission
from ongoing star formation and its associated gas supply within the
GRB\,080517 host galaxy.

\section{Discussion}\label{sec:disc}
Detections of GRB host galaxies in molecular gas are rare, at least in
part because of the difficulty of such observations, but the
remarkable feature of the host of GRB\,080517 is not its detection,
but rather that the emission is so weak. The inferred molecular gas
mass in the GRB host represents less than 20\% of the stellar mass
derived from its optical and near-infrared continuum flux. While this gas 
fraction is not, in itself, unusual, in combination with the galaxy's star
formation rate it presents an anomaly.

The on-going star formation
rate in the host galaxy of GRB\,080517 is
higher than typical for its stellar mass
 and suggests that it will burn through its available supply of
molecular gas in $\sim40$\,Myr, assuming 100\% conversion of gas to
stars. More conservative estimates for the efficiency of molecular gas
conversion only reduce this timescale. The short implied gas
consumption timescale suggests that the host of GRB\,080517 is
undergoing a short-lived star formation episode which is unlikely to
add significantly to its stellar mass. This conclusion is consistent
with the results of fitting the optical-near infrared spectral energy
distribution, which required that the ongoing starburst contributed
$<1$\% of the mass of the host galaxy, which is dominated by a more
mature, 500\,Myr-old stellar population \citep{2014arXiv1409.5791S}.

The width of the observed CO(1-0) emission line may lend tentative
support to such a scenario. If the line width, $\Delta v =
77$\,km\,s$^{-1}$, is interpreted as Doppler broadening due to the
stellar velocity dispersion, it implies a virial mass of just
$1.3\times10^{10}$\,M$_\odot$. This must include the SED-derived
stellar mass, 3.8$^{+0.2}_{-1.2}\times10^{9}$\,M$_\odot$ \citep{2014arXiv1409.5791S} and the gas
content. The fraction of molecular relative to atomic gas in GRB host
galaxies is poorly constrained, with existing measurements suggesting
$\sim$10\% \citep{2014arXiv1409.6315F,2013A&A...557A..18K} or lower
\citep{2014A&A...564A..38D} based on individual lines of sight probed
by ultraviolet spectra of GRB afterglows. Our estimated molecular gas mass
$M_\mathrm{H_2}\sim6.3\times10^8$\,$M_\odot$ would therefore imply an
atomic gas content similar to the galaxy's stellar mass. Comparison
with the virial mass suggests either that the host galaxy
has very little dark matter, or that the line emission is not fully sampling
the stellar velocity dispersion, as would be the case in a highly
inclined disk galaxy or if the emitting gas has been recently accreted
onto the GRB host and is not yet virialised.

Given the presence of a neighbouring galaxy, itself star-forming and
sufficiently close in both projection and velocity to constitute an
interacting system \citep{2014arXiv1409.5791S,2008AJ....135.1877E}, it
is tempting to speculate that the gas supply was accreted during a
recent near fly-by of the galaxy pair. However, we caution that the
current data has neither the signal to noise nor the spatial
resolution to identify tidal features or other direct evidence of
gravitational interaction. Alternate explanations remain
plausible. There is evidence that the CO(1-0) transition may be
sub-thermally excited in some galaxies with modest star formation
rates \citep[e.g.][]{2014arXiv1409.8158D} at $z>1$.  These are
typically higher in redshift and specific star formation rate and
lower in metallicity than GRB Host 080517
\citep{2014arXiv1409.5791S}. 
Measurement of
further rotational emission lines will be required to determine an
accurate spectral line excitation ladder, and hence accurate gas
temperature and mass.

Star formation in the local Universe is usually seen in relatively low
mass galaxies compared to those at higher redshifts - the well-known
`downsizing' phenomenon \citep{2004ApJ...603L..69C}. Where star formation is observed in massive
galaxies, it is typically accompanied by large molecular gas
reserves. As a result the timescale for depletion of molecular gas,
$\tau=$M(H$_2$)/SFR in years, scales with the stellar mass of a galaxy
\citep{2014arXiv1409.4764B}. 

\begin{figure*}
\epsscale{0.6}
\plotone{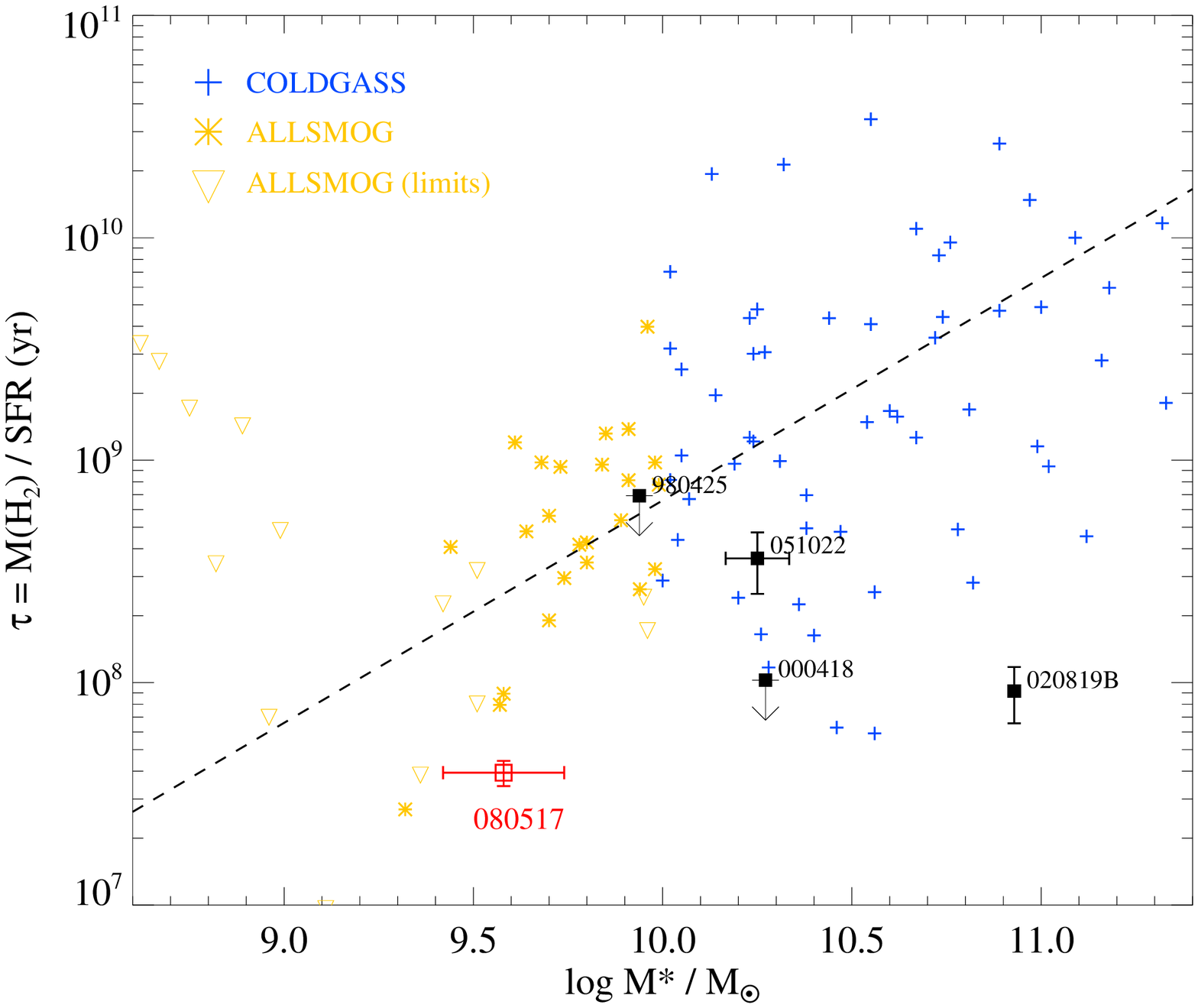}
\plotone{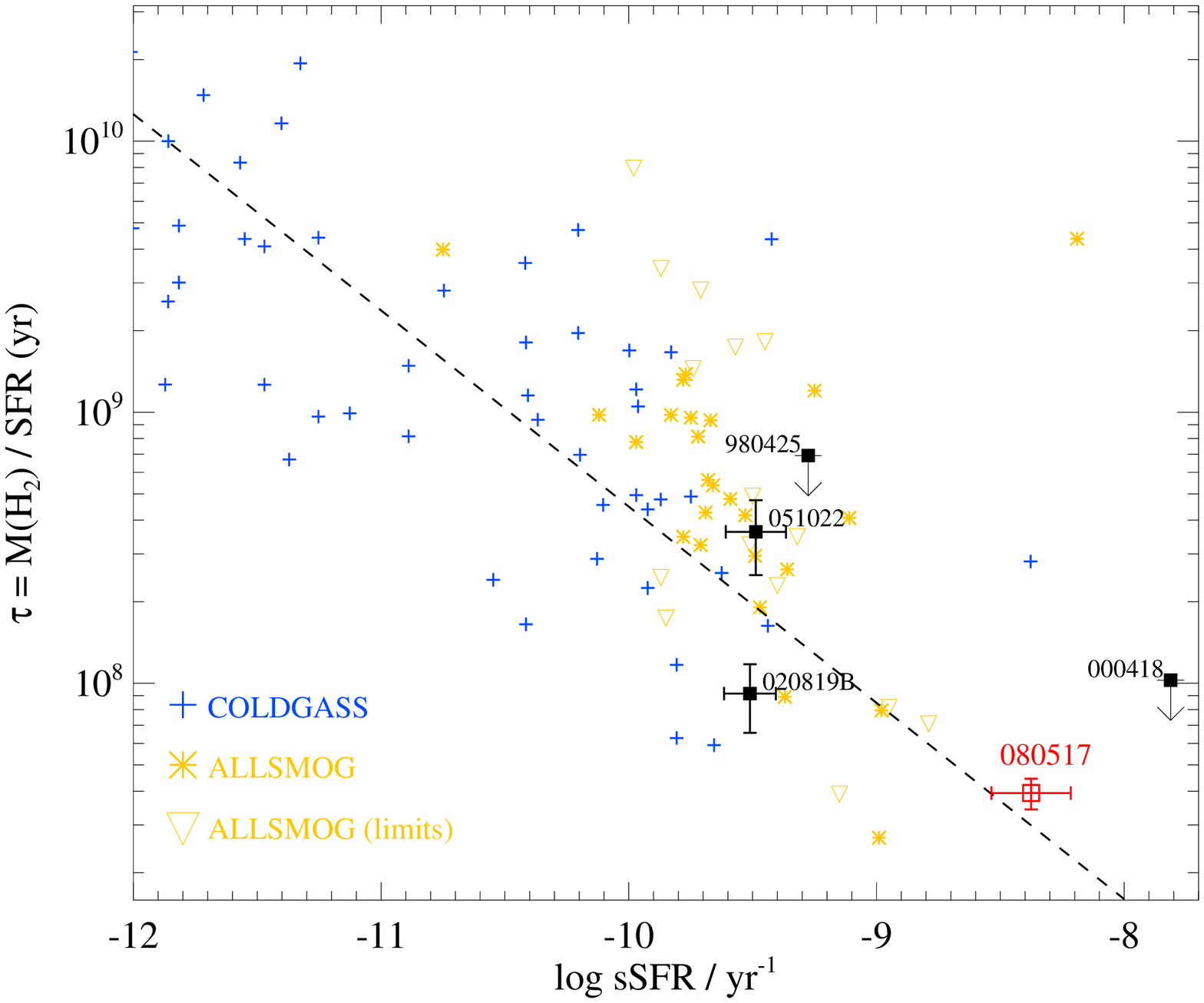}
\caption{The timescale for molecular gas consumption (in years) for GRB Host 080517, compared to local galaxies from the COLDGASS \citep{2011MNRAS.415...32S} and ALLSMOG \citep{2014arXiv1409.4764B} surveys. The dashed line marks the linear relation between stellar mass and depletion timescale determined by \citet{2014arXiv1409.4764B} for those samples, and the relation between timescale and specific star formation rate determined by \citet{2011MNRAS.415...61S}. We correct all literature values to our assumed CO-H$_2$ conversion factor, $\alpha$=4.3. Error bars combine 1\,$\sigma$ uncertainties in all derived values. In common with other observed GRB host galaxies, GRB Host 080517 lies well below the mean gas depletion timescale in the local Universe for its mass, but is on the expected relation for its specific star formation rate.\label{fig:depletion}}
\end{figure*}

As figure \ref{fig:depletion} demonstrates, the estimated gas mass and
star formation rate places the host galaxy of GRB\,080517 on the
established gas-to-specific star formation rate relation (main
sequence) for star forming galaxies in the local Universe, but well
away from the predicted gas consumption timescale for its mass
\citep{2011MNRAS.415...61S,2014arXiv1409.4764B}. For comparison, we
also mark the locations in this parameter space of those burst hosts
with comparable CO measurements
\citep{2014Natur.510..247H,2011ApJ...738...33H,2007ApJ...659.1431E},
drawing stellar mass and star formation rate estimates from the
literature \citep{2013ApJ...778..128P,2009ApJ...693..347M} where
necessary. We exclude the host of $z\sim8$ burst GRB\,090423 since it
remains undetected in the continuum
\citep{2014arXiv1408.2520B,2012ApJ...754...46T}, as well as in
molecular gas emission \citep{2011MNRAS.410.1496S}.

The host of GRB\,080517 appears to be typical of those GRB hosts
observed to date. These all show short gas depletion timescales, below
the relation derived by \citet{2014arXiv1409.4764B} and
\citet{2011MNRAS.415...61S} for star forming galaxies in the local
Universe. This is somewhat puzzling as a straightforward
interpretation would imply that GRBs typically occur towards the end
of a star formation episode, when little molecular gas remains. Such
an interpretation contrasts with the assumption that they arise from
massive stars, which collapse with only a short delay after the onset
of star formation. An alternate plausible scenario is that the starburst
giving rise to the GRB in these sources is a short-lived `flash in the pan' event,
not contributing substantially to the galaxy mass, as would seem to be the case
for GRB\,080517. This too would be a little surprising, since it implies that
all the bursts observed in molecular gas to date fall into this category with the bulk
properties of the host galaxy not representative of the brief star formation episode
underlying the burst.

We note that our analysis represents an independent confirmation of
the molecular gas deficit seen in absorption line studies of GRB
afterglows \citep[see][]{2007ApJ...668..667T,2009A&A...506..661L}. It
has been suggested that may be due in part to the low metallicity
typical of burst hosts \citep{2007ApJ...668..667T,2009A&A...506..661L}
with resultant low dust content, or that detection of H$_2$ absorption
requires significant depletion of refractory elements onto dust grains
\citep{2013A&A...557A..18K}, although the small spatial scales probed
by absorption studies along a line of sight complicate these
interpretations \citep{2014arXiv1409.6315F}. It seems unlikely that
metallicity is a strong factor in the short gas consumption timescales
seen in figure \ref{fig:depletion}. While all five bursts appear
consistent with shorter timescales than typical for their mass, GRBs
980425 and 000418 are believed to be sub-Solar in mean
metallicity, while GRBs\,020819B, 051022 and 080517 are likely Solar or super-Solar
\citep[see][]{2014arXiv1409.5791S,2014Natur.510..247H,2010AJ....139..694L,2010MNRAS.tmp..479S}.

On the other hand, the role of depletion of metals onto dust grains
may be significant. Of the bursts with CO detections, or deep limits,
to date; GRBs\, 051022 and 020819B are classified as `dark' bursts,
their optical afterglows likely sub-luminous due to dust extinction
\citep[see][]{2013ApJ...778..128P}, and GRBs\,080517 and 000418 show
evidence for dusty conditions in their host galaxy
\citep[080517,][]{2014arXiv1409.5791S} or afterglow
\citep[000418,][]{2000ApJ...545..271K}. The host of the low redshift,
low luminosity GRB\,980425, by contrast, is characterized by a low
dust content overall, but with a high density environment associated
with the GRB site \citep{2014A&A...562A..70M}. All five GRB hosts
investigated to date, therefore, have dust properties somewhat
atypical of the GRB population as a whole \citep[see e.g.][]{2010MNRAS.tmp..479S}.  Characterizing any
association between extinction and gas depletion timescales will
require larger, more complete samples. Obtaining these may be possible
in the near future, utilizing the new sensitivity of ALMA for southern
hemisphere targets. However, our observations with the (5-element)
Plateau de Bure Interferometer demonstrates that detections of low
redshift GRB hosts are possible with more moderate instrumentation.

\section{Conclusions}\label{sec:conc}

We have identified only the third long GRB host galaxy to show
observed emission from molecular gas. The integrated emission line
flux of GRB host 080517, $S\Delta\nu = 0.39\pm0.05$\,Jy\,km\,s$^{-1}$, suggests an
estimated molecular mass of
$M_\mathrm{H_2}\sim6.3\times10^8$\,$M_\odot$. This leads to a
remarkably short timescale for gas consumption in this system
($\sim$40\,Myr) and, together with constraints from other wavelengths,
suggest that the GRB occurred in a short-lived star formation episode
that does not dominate the galaxy's mass. 

It appears that the GRB host galaxies observed to date show lower gas masses
than might be anticipated from their mass and star formation rate, implying that
building this sample in future may be difficult. While GRB\,080517 is too far north for
follow-up observations with ALMA, we have demonstrated that it is accessible with older,
northern hemisphere arrays and that similar galaxies may be straightforwardly probed with
current instrumentation.

\acknowledgments
AJL and ERS are funded in part by STFC grant ST/L000733/1.
The research leading to these results has received funding from the European Commission Seventh Framework Programme (FP/2007-2013) under grant agreement No 283393 (RadioNet3).
We are very grateful to the PdBI staff observers and schedulers who secured good $uv$-plane coverage in this short integration.

{\it Facilities:} \facility{PdBI}

\end{document}